\documentclass[twocolumn,aps]{revtex4}
\input{epsfig.sty}
\newcommand{\beq}{\begin{eqnarray}}
\newcommand{\eeq}{\end{eqnarray}}
\begin{document}
\title{Electromagnetic Field Creation During EWPT Nucleation With Lepton Currents}
\author{Leonard S. Kisslinger and Sameer Walawalkar\\
   Department of Physics, Carnegie-Mellon University, Pittsburgh, PA 15213\\
  Ernest M. Henley\\
Department of Physics, University of Washington, Seattle, WA 98195\\
  Mikkel B. Johnson \\
Los Alamos National Laboratory, Los Alamos, NM 87545}
\begin{abstract}
   We include the electromagnetic currents from fermion degrees of freedom
in the equations of motion for electroweak MSSM with a right-handed
Stop that we have recently investigated. It is found that near the surface
of the bubble walls there are important effects on the electromagnetic
fields produced during bubble nucleation.
\end{abstract}
\maketitle
\noindent
PACS Indices:12.38.Lg,12.38.Mh,98.80.Cq,98.80Hw
\vspace{3mm}
\noindent

Keywords: Cosmology; Electroweak Phase Transition; Bubble Nucleation

   In the Standard Model, with the electromagnetic (em) field, the $W^{\pm}$
 and $Z$ fields, the gauge fields, a Higgs, and the standard leptons and 
quarks, there is no first order electroweak phase transition (EWPT)\cite{klrs}.
With a minimal supersymmetric extension of the standard model (MSSM) having
a Stop with a mass similar to the Higgs mass there can be a first order 
EWPT\cite{laine,cline,losada,bodeker}. Recemtly we have used such a MSSM
to derive the general equations of motion (e.o.m.), and with an I-spin ansatz
have estimated the electromagnetic field created during the nucleation of
the bubbles containing MSSM fields with a finite Higgs correlator and mass,
within the universe with a massless Higgs\cite{hjk05}. In that work fermions
were neglected, and the electromagnetic currents leading to the electromagnetic
fields arose solely from the charged W's. In the present note we estimate the
additional effect of electron currents on the production of electromagnetic 
fields during EWPT nucleation.

   In the present note we derive the e.o.m. of the MSSM with a Stop for the 
complete Lagrangian, including leptons. From this we derive the electromagnetic
field with the I-spin ansatz in order to determine the importance of the
leptonic currents. Since our objective is derive the magnetic seeds which
might explain the large galactic and intergalactic magnetic fields 
which have been observed\cite{gr}, and nucleation cannot produce
magnetic seeds due to spherical symmetry, the present study is a preliminary
one to find the initial conditions for collisions, during which magnetic
fields are produced. For this reason our numerical results are based on
electron currents, and we neglect the other charged leptons and quarks.
\vspace{3mm}

  The Lagrangian is
\beq
\label{1}
  {\cal L}^{MSSM} & = & {\cal L}^{1} + {\cal L}^{2}  + {\cal L}^{3}
+ {\cal L}^{lep}
\nonumber  \\
         {\cal L}^{1} & = & -\frac{1}{4}W^i_{\mu\nu}W^{i\mu\nu}
  -\frac{1}{4} B_{\mu\nu}B^{\mu\nu} \\ \nonumber
 {\cal L}^{2} & = & |(i\partial_{\mu} -\frac{g}{2} \tau \cdot W_\mu
 - \frac{g'}{2}B_\mu)\Phi|^2  -V(\Phi) \nonumber \\
 {\cal L}^{3} &=& |(i\partial_{\mu} -\frac{g_s}{2} \lambda^a C^a_\mu)\Phi_s|^2
    -V_{hs}(\Phi_s,\Phi) \nonumber \, ,
\eeq
where the pure $C^a_\mu$ term is omitted in ${\cal L}^{1}$ and  
\beq
\label{wmunu}
  W^i_{\mu\nu} & = & \partial_\mu W^i_\nu - \partial_\nu W^i_\mu
 - g \epsilon_{ijk} W^j_\mu W^k_\nu\\ \nonumber
 B_{\mu\nu} & = & \partial_\mu B_\nu -  \partial_\nu B_\mu \, ,
\eeq
where the $W^i$, with i = (1,2), are the $W^+,W^-$ fields, $C^a_\mu$
is an SU(3) gauge field, ($\Phi$, $\Phi_s$) are the (Higgs, right-handed 
Stop fields), $(\tau^i,\lambda^a)$ are the (SU(2),SU(3) generators, and
the electromagnetic field is defined as 
\beq
\label{2}
   A^{em}_\mu &=& \frac{1}{\sqrt{g^2 +g^{'2}}}(g'W^3_\mu +g B_\mu) \; .
\eeq
The leptonic Lagrangian is
\beq
\label{lep}
      {\cal L}^{lep} & = & \bar{\chi}_L\gamma^\mu(i\partial_{\mu} -\frac{g}{2} 
\tau \cdot W_\mu  - g'\frac{Y}{2}B_\mu) \chi_L \\ \nonumber
      && +\bar{\chi}_R\gamma^\mu(i\partial_{\mu} -g'\frac{Y}{2}B_\mu) \chi_R \\
      && + {\rm \;mass\;\;terms} \nonumber \; .
\eeq
The electric charge is given by $Q=\tau^3 + Y/2$, with $\tau^3 =-1/2 $ and
$Y=-1$ for the left-handed electron, $L$, and $\tau^3 =0 $ and $Y=-2$
for the right-handed electron, R. We only consider electrons in the present
work, neglecting the other leptons and the quarks, and the antifermions, which
would tend to move in the opposite direction to the fermions.
We also neglect lepton masses. The present note is just to explore the possible
importance of fermion currents in creating magnetic fields in EWPT bubble
nucleation and collisions. Parameters, other definitions and details are given 
in Ref \cite{hjk05}.

\begin{figure}
\begin{center}
\epsfig{file=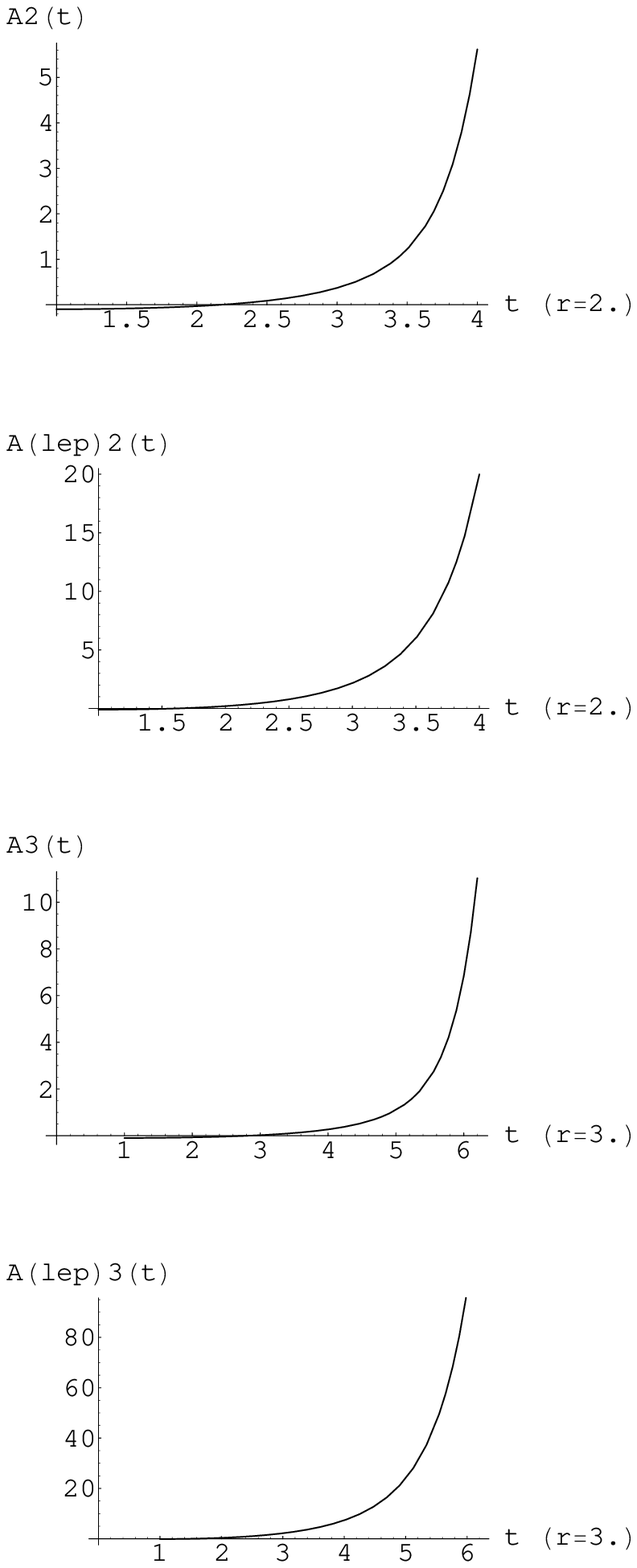,height=15cm,width=6cm}
\caption{A(t) without and A(lep)(t) with lepton currents for r=2.0 and 3.0}
{\label{Fig.1}}
\end{center}
\end{figure}

\begin{figure}
\begin{center}
\epsfig{file=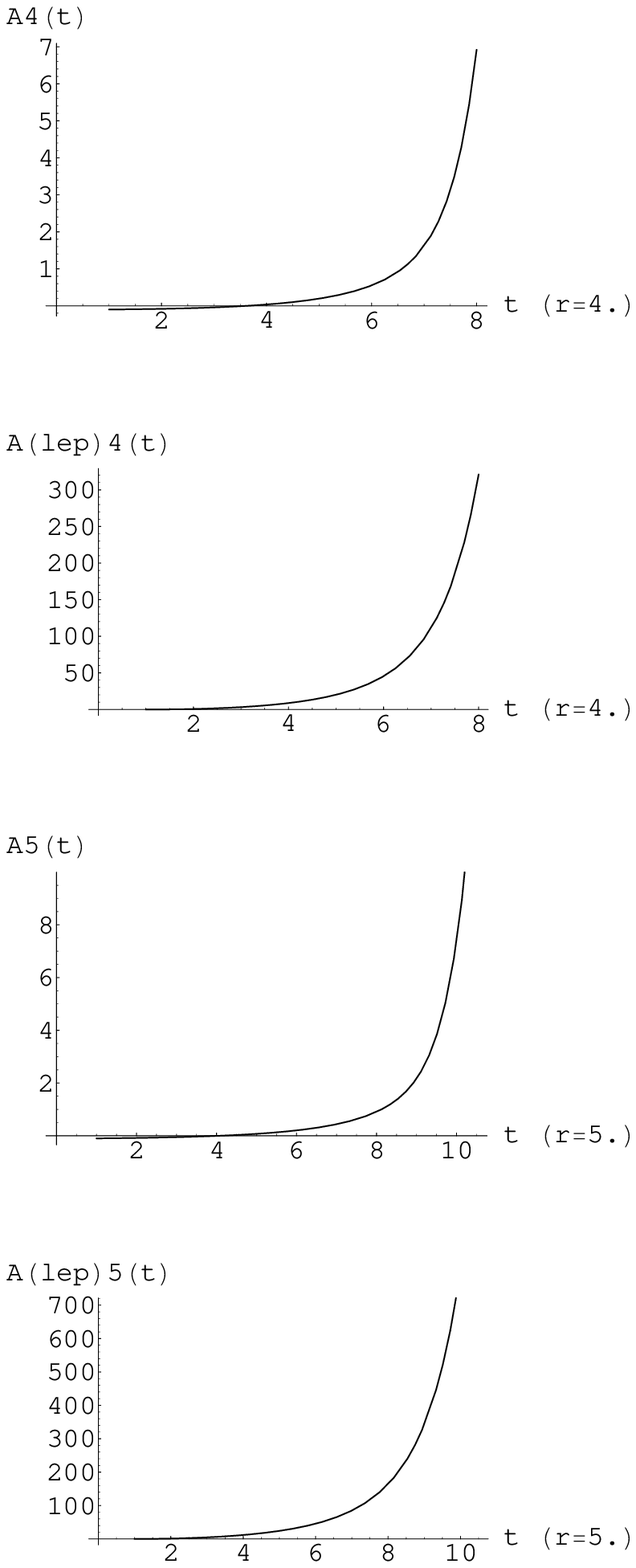,height=15cm,width=6cm}
\caption{A(t) without and A(lep)(t) with lepton currents for r=4.0 and 5.0}
{\label{Fig.2}}
\end{center}
\end{figure}

  The two e.o.m. needed for the present research are those for 
$W^{1}$, i=(1,2),and $A^{em}$
\beq
\label{eom1}
  && \partial^2 W^i_\nu-\partial^\mu \partial_\nu  W^i_\mu
 -g \epsilon^{ijk} {\cal W}^{jk}_\nu
 +\frac{g^2}{2}\rho(x)^2 W^i_\nu \nonumber \\
     && \;\;\; = 0 \; ,
\eeq
and
\beq
\label{eom2}
 &&\partial^2 A^{em}_\nu-\partial_\mu \partial_\nu  A^{em}_\mu
 -\frac{gg'}{\sqrt{g^2 + g^{'2}}} \epsilon^{3jk} {\cal W}^{jk}_\nu \nonumber \\
 &&\;\;\;\;\;-J^{lep}_\nu = 0 \; ,
\eeq
with
\beq
\label{3}
     {\cal W}^{jk}_\nu &\equiv&  \partial^\mu( W^j_\mu) W^k_\nu  
     + W^j_\mu \partial^\mu  W^k_\nu + W^{j\mu} W^k_{\mu\nu}\;, \nonumber \\
     W^k_{\mu\nu}&=& \partial^\mu W^k_\nu-\partial^\nu W^k_\mu-g\epsilon^{klm}
  W^l_\mu W^m_\nu \; ,
\eeq
and
\beq
\label{4}
       |\phi(x)|^2 &=& \rho(x)^2 \; .
\eeq
We have omitted the fermion contributions to the e.o.m. for $W^{1}$, 
Eq.(\ref{eom1}), since in the present study we are investigating the change
in $A^{em}$ due to the addition of the electron current in comparison to our
previous result.

For electron at finite temperature the leptonic current is\cite{kt}
\beq
\label{5}
     J^{lep}_\nu &=& G n_e \bar{u}_e \gamma_\nu u_e  \nonumber \\
              n_e &=& \frac{3}{4\pi^2} \zeta(3)T^3 \; ,
\eeq
with $ \zeta(3)\simeq 1.2$.
\vspace{3mm}

   As in Ref \cite{hjk05} we introduce the I-spin ansatz
\beq
\label{6}
  W^j_\nu &\simeq& i\tau^j W_\nu(x) \simeq i\tau^j x_\nu W(x)\;\; {\rm j=1,2,3}
 \nonumber \\
  A^{em}_\nu &\simeq& i\tau^3 A_\nu(x)\simeq i\tau^3 x_\nu A(x) \; ,
\eeq
with the I-spin operators defined as $\epsilon^{mjk} \tau^j\tau^k = i\tau^m$.
We use the Coulomb gauge
\beq
\label{7}
             \nabla \cdot \vec{W}(r,t)&=&\nabla \cdot \vec{A}(r,t)=0 \nonumber \\
                    W(r,t)&=& \frac{W(t)}{r^3} \\
                    A(r,t)&=& \frac{A(t)}{r^3} \; .\nonumber
\eeq

Eq(\ref{eom1}) for $\nu=t$ and $\nu=i=1,2$ becomes
\beq
\label{8}
   &&\frac{t}{r^3}(-\frac{6}{r^2}W(t)+\frac{gW(t)}{r^3}[(2t+\frac{r^2}{t})W'(t)
 -4W(t) \nonumber \\
            && +g\frac{r^2-t^2}{r^3}W(t)^2]+\frac{g^2}{2}\rho^2 W(t)) = 0 \\
\label{9}
  &&\frac{x_i}{r^3}(W''(t) -\frac{3}{r^2}W'(t)-\frac{3}{r^2}W(t)
+ \frac{gW(t)}{r^3}[3 t W'(t) \nonumber \\
          && -(1+\frac{3 t^2}{r^2})W(t) +\frac{g^2}{2}\rho^2 W(t))  =0 \; .
\eeq

\begin{figure}
\begin{center}
\epsfig{file=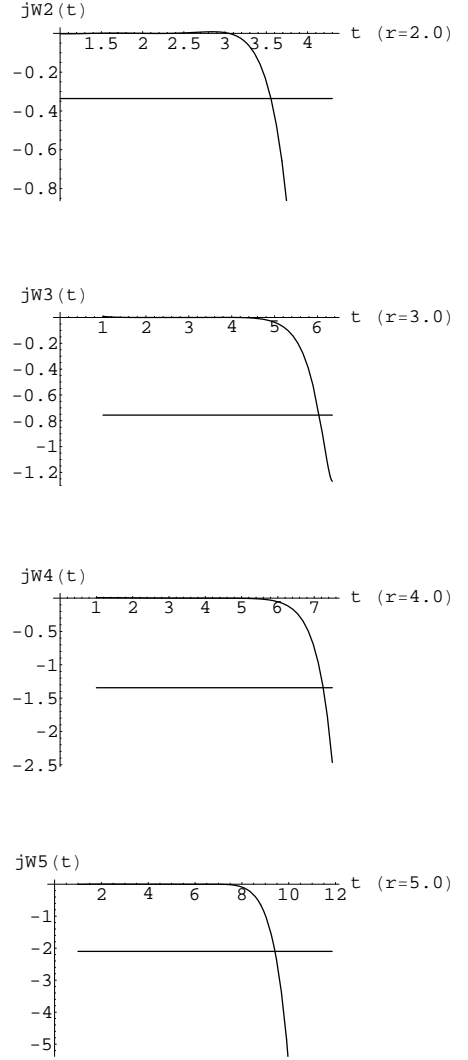,height=15cm,width=6cm}
\caption{jWr=W current for r=2.0, 3.0, 4.0, and 5.0 The horizontal lines are the
electron current = -0.084 r$^2$}
{\label{Fig.3}}
\end{center}
\end{figure}

Combining $r^3 (Eq(\ref{9})/x_i-Eq(\ref{8})/t)$ we find the e.o.m. for $W(t)$:
\beq
\label{eom3}
       W''(t)-\frac{3 t}{r^2}W'(t) +\frac{3}{r^2}W(t) +g H(t) &=& 0\; ,
\eeq
with
\beq
\label{10}
   H(t) &=& \frac{t^2-r^2}{r^3}W(t)[\frac{1}{t} W'(t)-\frac{3}{r^2}W(t)]\; .
\eeq

   Carrying out analogous calculations for $ A^{em}_\nu$ we find the e.o.m.
for $A(t)$
\beq
\label{eom4}
      A''-\frac{3 t}{r^2}A'+\frac{3}{r^2}A +G H -3 r^2 j &=& 0\; ,
\eeq
with $j=0.028 M_{Higgs}^3$. We have taken $T=M_{Higgs}$ at the time of
the EWPT.
\vspace{3mm}

Eqs(\ref{eom3},\ref{eom4}) are the same as those in Ref \cite{hjk05}
except for the leptonic current term in Eq(\ref{eom4}).
The solutions for Ar(t), by which we mean A(t) for fixed r, are shown in
Figs 1 and 2.

As one can see, the electron current plays an important role
in electromagnetic field production during nucleation of EWPT bubbles.
Although the electron current is much smaller than the current from
the $W^+,W^-$ fields near the bubble surface,  the electron
current is larger than the $W^+,W^-$ currents in the interior of the
bubble, as shown in Fig.3. The function $jWr(t)$ should be compared to
the electron current = -$0.084 r^2$.
This has an important effect on the complete soution.
Therefore, we conclude that the leptonic currents must be included
for the correct derivation of the magnetic fields during EWPT bubble
collisions.

\Large{{\bf Acknowledgements}}\\
\normalsize
This work was supported in part by the DOE contracts W-7405-ENG-36 and 
DE-FG02-97ER41014.

\end{document}